\documentstyle[aps,prl,multicol,graphicx]{revtex}
\begin{document}

\widetext
\title{Quantum Hall plateau transitions in disordered superconductors}

\author{V. Kagalovsky$^{1,2}$, B. Horovitz$^1$, Y.Avishai$^1$}
\address{$^1$Department of Physics, Ben-Gurion University of the Negev, 
Beer-Sheva 84105, Israel\\ 
$^2$The Technological College of Beer-Sheva,
School of Engineering, Beer-Sheva 84100, Israel}
\author{J. T. Chalker}
\address{Theoretical Physics, Oxford University, 
Oxford OX1 3NP, United Kingdom}

\maketitle

\begin{abstract}
We study a delocalization transition for non-interacting quasiparticles moving
in two dimensions, which belongs to a new symmetry class. This symmetry class
can be realised in a dirty, gapless superconductor in which time reversal
symmetry for orbital motion is broken,
but spin rotation symmetry
is intact. We find a direct transition between two insulating phases with
quantized Hall conductances of zero and two for the conserved quasiparticles.
The energy of quasiparticles
acts as a relevant, symmetry-breaking field 
at the critical point, which
splits the direct transition into two conventional 
plateau transitions.
\end{abstract}

\begin{multicols}{2}
The variety of universality classes possible in 
single-particle models of disordered conductors is now appreciated
to be quite rich. Three of these classes 
were identified early in the development of 
weak localization theory \cite{review}: they are
distinguished by the behavior of the system under time-reversal,
and by its spin properties, 
and are termed orthogonal, unitary and symplectic, in
analogy with Dyson's classification of random matrice ensembles.
Further alternatives can arise by two different 
mechanisms. First, in certain 
contexts, most notably
the integer quantum Hall effect, 
the non-linear $\sigma$ model describing a two-dimensional system
may admit a topological term \cite{Pruisken}, which results in the existence 
of extended states
at isolated energies in an otherwise localized spectrum.
Physically, such systems have more than one distinct insulating phase,
each characterised by its number of edge states, 
and separated from other phases by
delocalization transitions. 
Second, it may happen that the Hamiltonian
has an additional, discrete symmetry,
absent from Dyson's classification.
This is the case in two-sublattice
models for localization, if the Hamiltonian
has no matrix elements connecting states that belong to the same
sublattice \cite{gade,Hikami2}. It is also true
of the Bogoliubov-de Gennes formalism
for quasiparticles in a superconductor with disorder \cite{oppermann,AZ}. 
One consequence of this
extra symmetry is that,
at a delocalization transition, critical behavior can appear not only in
two-particle properties such as the conductivity, but also in
single-particle quantities, such as the density of states. 

Universality classes in systems with
extra discrete symmetries of this kind have attracted 
considerable attention from various directions. A general classification,
systematizing earlier discussions \cite{gade,oppermann},
has been set out by Altland and Zirnbauer \cite{AZ},
who examined mesoscopic normal-superconducting systems
as zero-dimensional realisations of some examples. Very recently,
quasiparticle transport
and weak localization has been studied in disordered, gapless
superconductors in higher dimensions,
with applications to normal-metal/superconductor junctions
\cite{simons}, and to thermal and spin
conductivity in high temperature superconductors 
\cite{zirnbauer,senthil,senthil2}.
Separately,
the behaviour of massless Dirac fermions in two space dimensions,
scattered by particle-hole symmetric disorder in the form
of a random vector potential, has been investigated intensively
\cite{dirac}
as a tractable example of a disordered critical point. 
And much before this, the one-dimensional tight binding model
with random nearest-neighbour hopping was shown \cite{1d}
to have a delocalization transition and divergent density of states
at the band center, the energy invariant under the sublattice symmetry.

In this paper, we study a new delocalization transition in two dimensions
that combines both of the above features: 
the transition separates phases
with  different quantized Hall conductances
for the quasiparticles, 
and it occurs in a system which
has a discrete microscopic symmetry.
This transition  can take place in a gapless superconductor
under appropriate conditions:
time-reversal invariance for orbital motion 
must be broken by an applied magnetic field, but the Zeeman coupling
should be negligible, so that the full spin-rotation invariance 
remains intact.
Formally, we suggest that the model
whose behavior we examine numerically
is a representative of the symmetry class labeled $C$ by
Altland and Zirnbauer \cite{AZ}, and that the delocalization transition is
associated with a topological term allowed in the field theory, as
noted by Senthil {\it et al} \cite{senthil}.
The possibility of
quantum Hall states in superconductors
with broken parity and time-reversal symmetry
has been emphasised by Laughlin \cite{laughlin}.
A direct transition into such a phase, in the presence of disorder,
is of particular interest in connection
with the theory of the quantum Hall effect, since it
is between phases with Hall conductance differing by two units.
Changes of Hall conductance by more than one unit
at a delocalization transition are 
precluded in generic systems, 
by the standard scaling flow diagram \cite{Khmelnitskii}
for the integer quantum 
Hall effect, and are possible in the model we study only because of 
additional symmetry.
In the presence of a symmetry-breaking coupling in the model,
of strength $\Delta$,
the transition occurs in two stages with a separation
that varies as $\Delta^{\phi}$ for small $\Delta$,
where $\phi \approx 1.3$.
Such a coupling is introduced if one treats
quasiparticle motion at finite energy. 

The system we consider is formulated as a generalisation of the
network model \cite{Chalker} for the quantum Hall plateau transition. 
The original version of this model describes guiding center motion
of spin-polarised electrons within one Landau level of a
disordered, two-dimensional system in a magnetic field.
It therefore has broken time reversal symmetry for orbital motion, 
and contains no spin degree of freedom.
It is specified in terms of scattering or transfer matrices,
defined on links and nodes of a lattice.
The model can be generalised in various ways. Spin can be incorporated
by allowing two amplitudes to propagate on each link.
This has been done previously \cite{Lee1,Wang,KHA1}, 
with the intention of describing
a spin-degenerate Landau level in which the two spin states are
coupled by spin-orbit scattering. In that case, the random $U(1)$ phases
which characterise propagation on the links of the 
original model are replaced with random $U(2)$ matrices, mixing the
two spin states without any rotational symmetry.
In the work presented here, we choose instead random $SU(2)$ 
matrices, preserving spin rotational symmetry.

In detail, the transfer matrix associated with each link
of the model is an SU(2) spin rotation matrix of the form
\begin{equation}
{\bf U}=\left( \begin{array}{cc}
e^{i\delta_1}\sqrt{1-x} & 
-e^{i\delta_2}\sqrt{x} \\
e^{-i\delta_2}\sqrt{x} & e^{-i\delta_1}\sqrt{1-x} 
\end{array} \right) .
\label{umatr1}
\end{equation}
where $\delta_1,\delta_2, x$ are random.
The transfer matrix at the nodes is parameterized by $\epsilon \pm 
\frac{1}{2}\Delta$ so that the transmission probability 
for the two spin states is
$[1+\exp (-\pi (\epsilon \pm \frac{1}{2}\Delta)]^{-1}$
respectively.
The value of $\epsilon$ determines
the Hall conductance of the system,
as measured at short distances:
varying $\epsilon$ drives the model through the delocalization transition.
A non-zero value for $\Delta$ breaks spin-rotation invariance,
and will in fact change the universality class for the transition.
Collecting factors, the transfer matrix across one 
node and the links connected to it is a 
$4\times 4$ matrix of the form \cite{Lee1}
\begin{equation}
{\bf T}=\begin{array}{ccc}
\left( \begin{array}{cc}{\bf U_1} & 0  \\
 0 & {\bf U_2}
\end{array}
\right) &
\left( \begin{array}{cc}{\bf C} & {\bf S}  \\
 {\bf S} & {\bf C}
\end{array}
\right) &
\left( \begin{array}{cc}{\bf U_3} & 0  \\
 0 & {\bf U_4}
\end{array}
\right) 
\end{array}
\label{tmatr}
\end{equation}
where
\begin{eqnarray}
&&{\bf C}=\left( \begin{array}{cc}
\sqrt{1+\exp [-\pi(\epsilon-\frac{1}{2}\Delta)]} & 0 \\
0 & \sqrt{1+\exp [-\pi(\epsilon +\frac{1}{2}\Delta)]} \end{array}
\right)
\nonumber \\
&&{\bf S}=\left( \begin{array}{cc}
\exp [-\frac{\pi}{2}(\epsilon-\frac{1}{2}\Delta)] & 0
\\
0 & \exp [-\frac{\pi}{2}(\epsilon +\frac{1}{2}\Delta)]
\end{array}
\right) 
\label{nodematr}
\end{eqnarray}
and ${\bf U}_i$ are as given in Eq\,(\ref{umatr1}).
From these $4 \times 4$ transfer matrices, ${\bf T}$, one can build up
a larger transfer matrix, of size $2M_l \times 2M_l$, with $M_l$ even,
to describe scattering in one slice of a system of width $M_l$ links,
(which has $M\equiv 2M_l$ scattering channels) 
by using independent realisations of ${\bf T}$ as diagonal blocks
of the larger matrix.

Both the $4 \times 4$ transfer matrix, ${\bf T}$, 
and the larger ones derived from
it, which we denote here also by ${\bf T}$,
are invariant under an antiunitary symmetry operation
representing spin reversal. The corresponding operator is 
${\bf Q}={\openone}\otimes i\tau_y\cdot K$,
where the Pauli matrix $\tau_y$ acts on the two spin states propagating
along each link, ${\openone}$ is the $M_l\times M_l$ unit matrix, and
$K$ is complex conjugation. 
The symmetry, ${\bf Q}{\bf T}{\bf Q}^{-1}={\bf T}$, holds for $\Delta=0$ only;
it implies that the Lyapunov exponents of the transfer matrix product 
have a two-fold degeneracy at $\Delta = 0$, which we
exploit in the analysis of our simulations, as described below.

It is possible to relate this network model to a Hamiltonian, $H$,
following Ref\,\onlinecite{ho}, by constructing a unitary matrix 
which can be interpreted as the evolution operator for the system,
for a unit time-step. Taking the continuum limit, one obtains, in the
case of the original network model for a spin-polarised 
Landau level, a two-component Dirac Hamiltonian
with random mass, scalar potential and vector potential.
For the network model of current interest, we get instead a
four-component Dirac Hamiltonian of the form
\begin{equation}
H=(\sigma_x p_x + \sigma_z p_z + m \sigma_y)\otimes\openone + 
\openone\otimes {\bbox{ \alpha}}\cdot{\bbox{ \tau}}\,,
\label{dirac}
\end{equation}
where $\sigma_i$ and $\tau_i$ for $i=x,y,z$ are two 
copies of the Pauli matrices, $\openone$ is the $2\times 2$
unit matrix, $p_x$ and $p_z$ are the two components of the 
momentum operator in the plane of the system, the mass $m$ is
proportional to $\epsilon$, the distance from the critical point,
and the real, three-component vector, ${\bbox{ \alpha}}$, 
is a random function of
position. This Hamiltonian has the symmetry
${\bf Q} H {\bf Q}^{-1}=-H$ \cite{footnote}, 
which is the defining feature of the
class labelled $C$ by 
Altland and Zirnbauer \cite{AZ}. A non-zero value for $\Delta$
in the network model introduces an additional term, 
$H'= \Delta\, \sigma_y\otimes \tau_z$ into the Dirac Hamiltonian,
breaking the symmetry.
Equally, since the symmetry relates eigenstates with energies $\pm E$,
and leaves invariant only those at energy $E=0$,
non-zero $E$, like $\Delta$, acts as a symmetry-breaking perturbation.

The models represented by Eqs\,(\ref{tmatr}) and (\ref{dirac})
describe propagation of quasiparticles
which are conserved, and which are obtained within
the Bogoluibov-de Gennes formalism 
by making a particle-hole transformation on 
states with one spin orientation (see, for example, Ref\,\cite{AZ}).
Specifically, starting from the  Bogoluibov-de Gennes Hamiltonian
for a singlet superconductor
\begin{eqnarray}
\nonumber
H_{S}=\sum_{ij}(h_{ij}[c^{\dagger}_{i\uparrow}c_{j\uparrow} 
+c^{\dagger}_{i\downarrow}c_{j\downarrow}] + \Delta_{ij}c^{\dagger}_{i\uparrow}c^{\dagger}_{j\downarrow} +
\Delta^{*}_{ij}c_{j\downarrow}c_{i\uparrow})
\end{eqnarray}
and introducing transformed operators,
$\gamma_{i\uparrow}=c_{i\uparrow}$ and $\gamma^{\dagger}_{i\downarrow}=c_{i\downarrow}$,
one has
\begin{equation}
H_{S}=
\sum_{ij}
\left(\begin{array}{cc} \gamma^{\dagger}_{i\uparrow} & \gamma^{\dagger}_{i\downarrow}
\end{array}\right)
\left( \begin{array}{cc}  h_{ij}  &  \Delta_{ij} \\
 \Delta^{*} _{i,j}  & -h^{T}_{ij}
\end{array}\right)
\left( \begin{array}{c}\gamma_{j\uparrow}\\ \gamma_{j\downarrow} \end{array}
\right)
\end{equation}
This Hamiltonian,
like the $SU(2)$ network model that we simulate,
has the symmetry ${\bf Q}H_S{\bf Q}^{-1} = -H_S$
\cite{AZ}, where here ${\bf Q}=i \tau_y K$ and $\tau_y$ acts on the particle-hole spinor of Eq. (5). For
singlet pairing (maintaining spin-rotation invariance) $\Delta_{ij}$ is
symmetric while $h_{i,j}$ is hermitian and the symmetry under
${\bf Q}$ is obvious; 
time-reversal symmetry is broken if $h\not= h^*$ or $\Delta \not= \Delta^*$.
Since quasiparticle charge in a superconductor is not conserved, 
and charge response is anyway controlled by the condensate,
the localization problem of interest in a disordered superconductor, 
as emphasised in
Ref\,\onlinecite{senthil}, involves 
spin and energy transport, rather than charge transport.
We stress that the Hall conductance examined below
is a property of quasiparticles described by a Hamiltonian
such as Eq\,(\ref{dirac}).

We study the model defined from Eq\,(\ref{tmatr})
at a range of values for $\epsilon$ and $\Delta$.
Preliminary calculations, reported earlier \cite{KHA2},
were limited to $\Delta=0$.
We compute
the normalized   
localization length, $\xi_M/M_l$, for strips of width
$M=16, 32, 64$, with periodic boundary conditions.   
For small $\Delta$ we find it necessary also to use
$M=128, 256$, in order to identify more clearly the 
critical properties.
The parameters $\delta_1/2\pi,\delta_2/2\pi, x$ 
are each chosen to be uniformly   
distributed in the interval $[0,1]$,
so that each matrix $\bf U$ associated with a link
is independently distributed with the Haar measure on $SU(2)$.
Runs were carried out for
strips of 
length 60,000 (for $M=16, 32$), 240,000 (for $M=64$), and 480,000 
($M=128, 256$). 
The errors are typically 0.5\%, except for $M=256$.

The behavior at non-zero $\Delta$ is shown in Fig.\,1, region I.
For $\Delta=2.0$,
extended states ($\xi_M/M_l$ independent of $M$)
appear clearly at two energies, $\pm\epsilon_c$, with  
$\epsilon_c(\Delta=2) \approx 0.6$. A one-parameter 
scaling fit for $\xi_M/M=f[(\epsilon -\epsilon_ c)M^{1/\nu_0}]$
yields $\nu_0 \approx 2.5$, the conventional quantum Hall exponent 
\cite{Lee1,KHA2,huckestein}.
For $\Delta = 0.2$, $\epsilon_c$ is too small to be resolved by this method
(Fig.\,1, region II). 
Proceeding in this way at a range of values for $\Delta$,
we construct the phase diagram for the model, shown in Fig.\,2.
With $\Delta \not= 0$, the two phases at $\epsilon \ll -1$ and $\epsilon \gg +1$
are separated by an intermediate, small $\epsilon$, phase.
Counting edge states in each phase in the strong-coupling limit
($\epsilon \to \pm \infty$ at fixed $\Delta$, and
$\Delta \to \infty$ at $\epsilon = 0$, respectively),
we find that the quasiparticle Hall conductance
takes the values $0$, $1$, and $2$ in successive phases with 
increasing $\epsilon$. If $\Delta$ is made smaller, the boundaries 
of the intermediate phase approach each other, and at $\Delta = 0$
there appears to be a direct transition between phases with Hall
conductance differing by two units.

To study this direct transition, we examine behavior at small
$\Delta$ in more detail.
On the line $\Delta=0$, the localization length 
diverges at a single critical point, $\epsilon_c=0$,
with an exponent, $\nu = 1.12$, which is different from that at the
conventional plateau transition (Fig.\,3, curve I).
Close to this line, scaling with system size is quite complex
and, in particular, the variation of $\xi_M/M$ with $M$ is not monotonic.
In order to extract scaling properties at small, non-zero $\Delta$, 
we monitor the deviation from 
Kramer's degeneracy 
of the smallest two positive Lyapunov exponents, $\lambda_1$ and
$\lambda_2$, of the transfer matrix product that represents the sample,
defining ${\bar \xi}=M(\lambda_2 -\lambda_1)$. 
Finite size scaling of both ${\bar \xi}$ and ${\xi_M}$
is shown in Fig.\,3 (curves II and III), as a function of $\Delta$ along the symmetry line
$\epsilon=0$. We find  for deviations from Kramers degeneracy
${\bar \xi}=f_1(\Delta M_l^{1/\mu})$ 
and for the localization length 
$\xi_M/M=f_2(\Delta M_l^{1/\mu})$, with $\mu \approx 1.45$.
The maximum in $f_2$ indicates that states at
$\epsilon =0, \, \Delta \neq 0$ appear delocalized at small $M$ 
while at larger $M$ the true, large-distance localized nature
becomes apparent. Thus to 
see localization one needs $M\gtrsim (9/\Delta)^{1.45}$: for example, 
for $\Delta =0.2$ one needs $M\gtrsim  200$.

We propose, then, that $\epsilon=\Delta=0$ is a critical point at which 
$\epsilon$ parameterizes the symmetry-preserving
relevant direction, and $\Delta$ is a symmetry-breaking
field,
so that $\xi_M/M_l$ 
is described near the fixed point by a two parameter scaling function
\begin{equation}
\xi_M/M_l =f(\epsilon M_l^{1/\nu}, \Delta M_l^{1/\mu})\,.
\label{scaling}
\end{equation}
with $\nu=1.12$ and $\mu\approx 1.45$.
In the presence of a symmetry-breaking field, 
$\Delta \neq 0$, scaling flow is
away from the new fixed point, giving quantum Hall plateau phases
except on trajectories
which connect this unstable fixed point to fixed points
at finite $\Delta$, representing the conventional universality class for 
plateau transitions. At these, a finite critical $\epsilon _c (\Delta)$ is 
expected with an exponent $\nu_0$, so that $\xi \sim [\epsilon - 
\epsilon_c(\Delta)]^{-\nu_0}$. Since the values 
of $\epsilon M^{1/\nu}$ and
$\Delta M^{1/\mu}$ on a critical trajectory 
serve to define a one-parameter curve, we expect
$\epsilon_c(\Delta) = \pm c \Delta^{\mu/\nu}$.
Thus, as $\Delta $ approaches zero, extended states coalesce, having 
a separation, $2 \epsilon_c \propto \Delta^{1.3}$ (the line in Fig.\,2),
which is much smaller than $\Delta$, their separation 
in the absence of coupling between the two
spin orientations.

A further aspect of the critical point which is of interest,
but not accessible within our numerical approach, is the behavior of
single-particle quantities such as the density of states,
discussed recently in Ref\,\cite{senthil2}.
We expect for the Hamiltonian of Eq\,(\ref{dirac}) a finite
density of states at all energies provided $\Delta \not= 0$, and
singularities in the density of states at zero energy when $\Delta = 0$,
with a different nature according to whether
$\epsilon=0$ or $\epsilon \not= 0$.
 
In conclusion, we have shown that quantum Hall plateau transitions
belonging to a new universality class occur in a model
for a gapless superconductor which is invariant under spin rotations, but
which has time-reversal symmetry broken for orbital motion.
In contrast to the conventional plateau transition, the
Hall conductance for conserved quasiparticles
changes at this transition by two units.
We have examined critical behavior,
and shown that there is a symmetry-breaking perturbation
which is relevant at the critical point, splitting the
transition into two, with extended states that coalesce as
the symmetry-breaking field is removed.

\vspace{5mm}
{\bf Acknowledgments}: We thank O. Agam, N. Argaman, 
S. Hikami, B. Huckestein, D. E. Khmelnitskii 
and Y. Meir for stimulating discussions. One of us (JTC) is 
particularly grateful to T. Senthil for many helpful comments, 
and to the Dozor Fellowship Program for support.

\end{multicols}

\begin{figure}
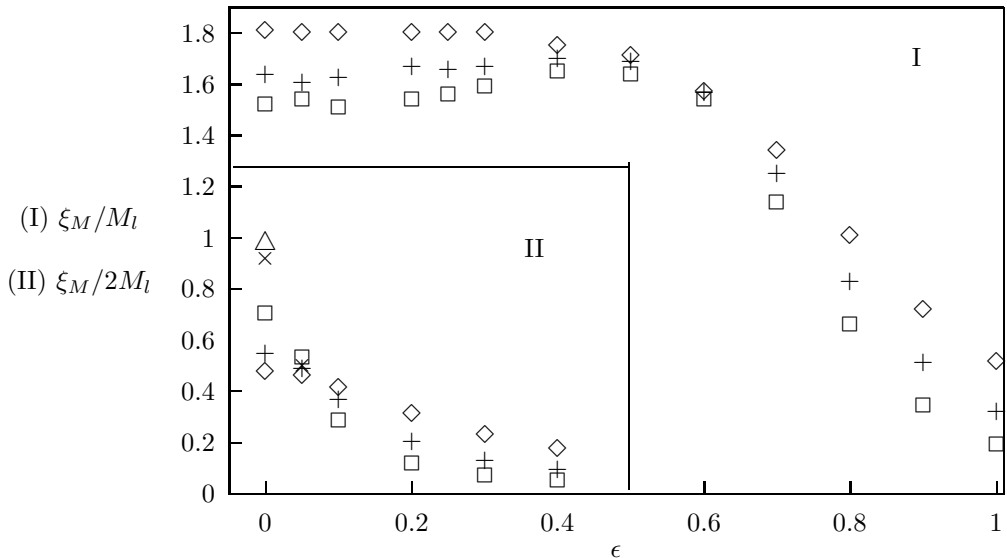

\setlength{\unitlength}{0.240900pt}
\ifx\plotpoint\undefined\newsavebox{\plotpoint}\fi
\sbox{\plotpoint}{\rule[-0.200pt]{0.400pt}{0.400pt}}%

\caption{Normalized localization length $\xi_M/M_l$.
The symbols $\Diamond$, $+$, $\Box$, $\times$, and $\triangle$ 
correspond to system widths $M$ of $16$, $32$, $64$, $128$, and $256$ respectively.
The symmetry breaking parameter is $\Delta=2$ (region I) or 
$\Delta=0.2$ (region II).}
\end{figure}

\begin{figure}
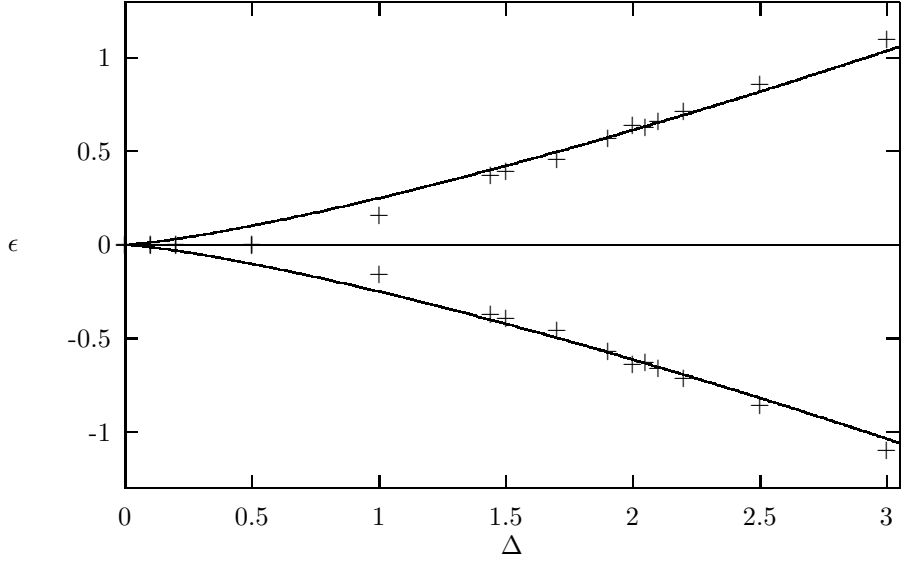

\setlength{\unitlength}{0.240900pt}
\ifx\plotpoint\undefined\newsavebox{\plotpoint}\fi
\sbox{\plotpoint}{\rule[-0.200pt]{0.400pt}{0.400pt}}%
%
\caption{Phase diagram. The $+$ symbols indicate fitted positions of extended states, and the lines are $\epsilon = \pm 0.25 \Delta^{1.3}$.}
\end{figure}

\begin{figure}
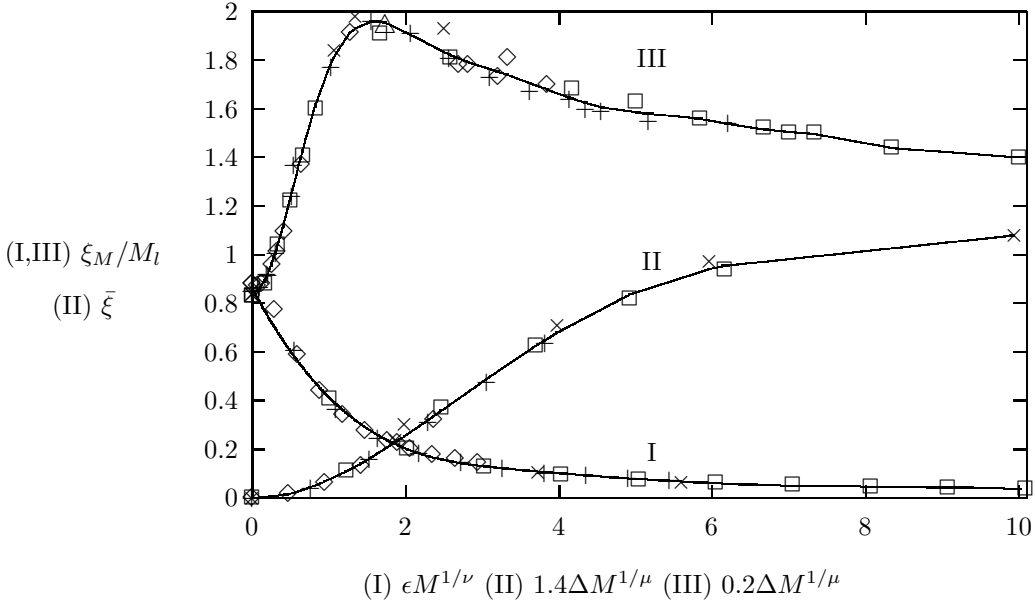

\setlength{\unitlength}{0.240900pt}
\ifx\plotpoint\undefined\newsavebox{\plotpoint}\fi
\sbox{\plotpoint}{\rule[-0.200pt]{0.400pt}{0.400pt}}%
%
\vspace{1cm}
\caption{Scaling functions:
(I) Normalized localization length $\xi_M/M_l$ as
function of $\epsilon M^{1/\nu}$ with $\nu=1.12$ for $\Delta =0$.
(II) Deviation from Kramer's degeneracy $\bar{\xi}$ as
function of $\Delta M^{1/\mu}$ with $\mu=1.45$ for $\epsilon =0$.
(III) $\xi_M/M_l$ as function of $\Delta M^{1/\mu}$ with $\mu=1.45$ 
for $\epsilon =0$.
Symbols denote system widths as in Fig.\,1.}

\end{figure}

\end{document}